\documentclass[aps,prb,twocolumn,superscriptaddress,showpacs]{revtex4}

\usepackage{graphicx}
\usepackage{amsmath}
\usepackage{amsmath}

\usepackage{amsmath}
\begin{document}

\title{Andreev bound states for superconducting-ferromagnetic box}

\author{J. Cserti}
\email{cserti@galahad.elte.hu}
\affiliation{Department of Physics of Complex Systems, E{\"o}tv{\"o}s
University}
\author{J. Koltai}
\affiliation{Department of Biological Physics, E{\"o}tv{\"o}s
University, H-1117 Budapest, P\'azm\'any P{\'e}ter s{\'e}t\'any 1/A,
Hungary}
\author{C.~J. Lambert}
\email{c.lambert@lancaster.ac.uk}
\affiliation{Department of Physics, Lancaster University, Lancaster,
LA1 4YB, UK}

%\date{\today}

\begin{abstract}
Within the microscopic Bogoliubov--de~Gennes (BdG) formalism an exact 
quantization condition for Andreev bound states of the 
ferromagnetic-superconducting hybrid systems of box geometry is derived 
and a semi-classical formula for the density of states is obtained. 
The semi-classical formula is shown to agree with the exact result, 
even when the exchange field $h$, is much larger than the superconductor 
order parameter, provided $h$ is small compared with the Fermi energy. 

\end{abstract}


\maketitle

Mesoscopic hybrid systems formed from ferromagnets (F) in contact
with superconductors (S) exhibit interesting transport properties 
resulting from the suppression of the electron-hole
correlation in the 
ferromagnets~\cite{deJong-Carlo,Fabio,Kikuchi,Zutic-1,Zutic-2,Belzig,Kadigrobov,Krawiec-Gyorffy}.  
These transport phenomena are intimately related to the
influence of an exchange field on the density of states (DOS) of 
clean ferromagnetic films in contact with superconductor, which has 
been investigated both experimentally\cite{Kontos} and 
theoretically\cite{Nazarov-Ferro}.  
In FS systems, Andreev bound states below the bulk 
superconducting gap are spin split by the exchange field of 
the ferromagnet.
In a quasi-classical treatment of the diffusive regime it was shown 
that sub-gap features in the DOS of FS hybrids 
can be understood from the behavior of the length distribution of the 
classical trajectories existing in the ferromagnetic region, which 
depends purely on the geometry and the boundaries of the
ferromagnetic region.

\begin{figure}[hbt]
\includegraphics[scale=0.5]{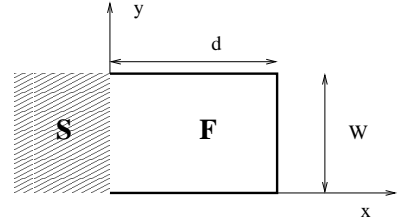}
\caption{A rectangular shape of ferromagnetic dot (F) in contact with a
superconductor (S).
\label{geometria-fig}}
\end{figure}

In this work we calculate the Andreev levels of an FS box  
consisting of a clean ferromagnetic region with a 
superconductor attached to one side, as 
shown in Fig.~\ref{geometria-fig}.
This geometrical arrangement is a generalization of that
investigated in Ref.~\onlinecite{Nazarov-Ferro}, 
where the size of the system along the FS interface was infinite. 
The discrete energy spectrum of the FS box system is obtained by solving 
the microscopic Bogoliubov-de Gennes (BdG) 
equation\cite{deJong-Carlo,BdG-eq-deGennes}. 
We derive an exact quantization condition without using the frequently 
applied Andreev's approximation. Our exact quantum description of the FS
hybrid is an extension of the commonly used model developed 
by Blonder et al.~\cite{BTK} and 
by Saint-James and de Gennes\cite{deGennes-Saint-James} 
for normal-superconducting systems.    
The mismatch in the effective masses and Fermi energies of the ferromagnet 
and the superconductor are taken into account in our calculations and the 
tunnel barrier at the interface is modeled using a Dirac delta potential.  
The treatment of the problem is based on an adaptation of a method 
developed in our previous work~\cite{box_disk:cikk} to the case of 
FS systems. 

Starting from this exact quantization condition, 
we give a semi-classical expression for the sub-gap density of states (DOS) 
for exchange fields which are much less than the Fermi energy. 
The semi-classical DOS is expressed in terms of the classical return 
probability of the electrons which depends only on the geometry of the 
F region and its boundaries.    
Besides the DOS,  from our derivation an explicit expression for this
return probability is also found.  
Based on the quasi-classical Green's function approach a similar 
expression has been derived for the DOS of ferromagnetic 
layers in contact with clean superconductor
in Ref.~\onlinecite{Nazarov-Ferro}.  
Our DOS expression for FS systems can be regarded as an extension of 
the Bohr-Sommerfeld formula developed for normal-superconducting 
hybrids\cite{BS-DOS}.
In this work we compare the DOS obtained from the exact quantum
calculations with that found from our semi-classical formula.
We show that a good agreement between the two calculations can be expected 
only for small enough exchange fields (much less than the Fermi energy). 

The BdG equation for the FS systems can be written as
\begin{equation}
\left(\begin{array}{cc}
H_0 - \sigma h({\bf r}) & \Delta \\ 
\Delta^* & - H_0 - \sigma h({\bf r})
\end{array}    
 \right)
\Psi_\sigma
= E \,  \Psi_\sigma,
\end{equation}
where $H_0 = {\bf p}^2/2m_{\rm eff} + V({\bf r})- \mu$ is 
the single-particle Hamiltonian,    
$\mu = E_{\rm F}^{(F)}, E_{\rm F}^{(S)}$ are the Fermi energies,  
$m_{\rm eff}= m_{\rm F}, m_{\rm S}$ are the effective masses 
in the F/S regions, $\Psi_\sigma$ is a two-component wave function,  
$E$ is the quasi-particle energy measured from the Fermi energy 
$E_{\rm F}^{(F)}$.
Here $\sigma = \pm 1$ for spin up/down states and $+/-$ refer to the
electron/hole like quasi-particle excitation. 
The tunnel barrier $V({\bf r})$ at the FS interface and 
the exchange energy $h({\bf r})$ are modeled in a usual way 
by $V(x,y)=U_0 \, \delta(x)$ and $h(x,y) = h \Theta (x)$, 
where $\Theta$ is the unit step function. 
We also adopt the usual step-function model\cite{deJong-Carlo,BTK} 
for the pair potential and take $\Delta({\bf r})= \Delta \Theta (x)$. 
It is easy to see that the Hamiltonian is separable and 
the ansatz for the wave functions in the F region can be written as 
\begin{subequations}
\label{hullfv:e}
\begin{equation}
\Psi_{m,\sigma} (x,y) =  
\left( \begin{array}{l} 
a_+ \sin \left[ k_{m,\sigma}^{\left(+\right)} 
\left( x-d\right)\right]  \\[1ex]
a_- \sin \left[ k_{m,\sigma}^{\left(-\right)} 
\left( x-d\right)\right]
\end{array} \right) \chi_m(y), 
\label{hullfvF}
\end{equation}
while in the S region the wave functions have the form 
\begin{equation}
\Psi_{m,\sigma} (x,y) = \left( \begin{array}{l} 
c_+ \gamma_+ \, e^{-iq_{m,\sigma}^{\left(+\right)}x } 
+ c_- \gamma_- \, e^{iq_{m,\sigma}^{\left(-\right)}x }   \\[1ex]
c_+ \, e^{-iq_{m,\sigma}^{\left(+\right)}x } 
+ c_- \, e^{iq_{m,\sigma}^{\left(-\right)}x } 
\end{array} \right) \chi_m(y), 
\label{hullfvS}
\end{equation}
where 
\begin{eqnarray}
k_{m,\sigma}^{\left(\pm\right)}  &=& k_{\rm{F}}^{\rm{(F)}}
\sqrt{1 \pm \frac{E+\sigma h}{E_{\rm{F}}^{\rm{(F)}}}
-{\left(\frac{m \pi}{k_{\rm{F}}^{\rm{(F)}}W}\right)}^2},   \\
q_{m,\sigma}^{\left(\pm\right)}  &=& 
k_{\rm{F}}^{\rm{(S)}}
\sqrt{1 \pm i\,\frac{\sqrt{E^2 - \Delta^2}}{E_{\rm{F}}^{\rm{(S)}}} 
-{\left(\frac{m \pi}{k_{\rm{F}}^{\rm{(S)}}W}\right)}^2}, \\
\chi_m(y) &=& \sqrt{2/W} \, \sin (m\pi y/W), \\[1ex]
\gamma_{\pm} &=& e^{\pm i \arccos \left(E/\Delta \right)}. 
\end{eqnarray}
\end{subequations}
Here $m$ is a fixed integer and the Fermi wave numbers 
in the F/S regions are given by $k_{\rm{F}}^{\rm{(F)}} = 
\sqrt{2 m_{\rm F}  E_{\rm{F}}^{\rm{(F)}}/ \hbar^2}$ and 
$k_{\rm{F}}^{\rm{(S)}} = 
\sqrt{2 m_{\rm S}  E_{\rm{F}}^{\rm{(S)}}/ \hbar^2}.$ 
The wave functions satisfy the Dirichlet boundary conditions 
at the boundary of the F region except for the FS interface where 
the matching conditions~\cite{box_disk:cikk} should be applied. 
The four coefficients $a_{\pm},  c_{\pm}$ in Eqs.~(\ref{hullfvF}) and
(\ref{hullfvS}) are determined from these matching conditions. 
One can find the following secular equation for the eigenvalues 
$E$  of the FS system for fixed mode index $m$ 
and spin state $\sigma$: 
\begin{subequations}
\begin{equation} 
\rm{Im} \left \{\gamma_+ D^{\rm{(+)}}_{m,\sigma}(E,h) \, 
D^{\rm{(-)}}_{m,\sigma}(E,h) \right \}=0,
\label{DNS}
\end{equation}
where ${\rm Im}\{. \}$ stands for the imaginary part and 
\begin{eqnarray}
D^{\rm{(+)}}_{m,\sigma}(E,h)  &=& 
\left(Z- i\frac{m_{\rm F}}{m_{\rm S}} q_{m,\sigma}^{\left(+\right)} \right)
\sin k_{m,\sigma}^{\left(+\right)} d  \nonumber \\
&& + k_{m,\sigma}^{\left(+\right)} \, \cos k_{m,\sigma}^{\left(+\right)}d, \\
D^{\rm{(-)}}_{m,\sigma}(E,h) &=&
\left(Z+ i\frac{m_{\rm F}}{m_{\rm S}} q_{m,\sigma}^{\left(-\right)} \right)
\sin k_{m,\sigma}^{\left(-\right)} d  \nonumber \\
 && + k_{m,\sigma}^{\left(-\right)} \, \cos k_{m,\sigma}^{\left(-\right)}d, 
%\left[D^{\rm{(+)}}_{m,\sigma}(-E,-h)\right]^{*}, 
\end{eqnarray}
\label{De}
\end{subequations}
and $Z=2m_{\rm F}U_0/\hbar^2$ is the normalized barrier strength. 
The number of propagating modes for the electron/hole are 
the maximum of $m$ for which $k_{m,\sigma}^{\left(\pm\right)}$ is a 
real number, i.e., 
$M_\sigma^{\left(\pm \right)}= 
\left[M\sqrt{1\pm\left(E+\sigma h \right)/E_F^{\left({\rm F}\right)}} 
\right]$, where $M=k_{\rm F}^{\rm{(F)}}W/\pi$ and $\left[\cdot \right]$ 
stands for the integer part. 
For non-propagating modes $k_{m,\sigma}^{\left(\pm\right)}$ have to be
replaced with their imaginary part and the functions $\sin$ and $\cos$ 
with the functions $\sinh$ and $\cosh$, respectively.
For fixed $m$ and $\sigma$ the solutions of Eq.~(\ref{DNS}) for $E$ give
the discrete sub-gap energy spectrum ($E< \Delta$). These levels are
numerically exact, i.e., no Andreev's approximation is used. 

We now calculate density of states below the gap. In what follows,
we assume that there is no mismatch and tunnel barrier at the FS
interface ($m_{\rm F} =m_{\rm S}$, $E_{\rm{F}}^{\rm{(F)}}
=E_{\rm{F}}^{\rm{(S)}}$ and $Z=0$). 
For simplicity, we shall omit the superscript F and S in the wave
numbers and the Fermi energies.   
In Andreev's approximation, i.e., 
for $\left| E+ \sigma h \right| \ll E_{\rm F}$ we have 
 $k_{m,\sigma} \approx q_{m,\sigma}$ and 
$D^{\rm{(+)}}_{m,\sigma}(E,h) \approx k_{m,\sigma}^{\left(+\right)} \, 
e^{-ik_{m,\sigma}^{\left(+\right)}\, d} $. Therefore, the quantization
condition (\ref{DNS}) can be simplified
\begin{equation}
I_{m,\sigma}(E) \equiv 
\frac{\left(k_{m,\sigma}^{\left(+\right)}  -k_{m,\sigma}^{\left(-\right)} 
\right) d 
- \arccos \left(E/\Delta \right)}{\pi} = n. 
\label{Fm:eq}
\end{equation}
The density of states for energies below the gap 
($\left|E \right| \le \Delta$~) is  
\begin{equation}
\varrho(E) =  \sum_{\sigma=\pm 1}\sum_{n=-\infty}^{\infty} \, 
\sum_{m=1}^{M_0}\, \delta(E - E_{mn,\sigma}),  
\end{equation}  
where $E_{mn,\sigma}$ are the solutions of Eq.~(\ref{Fm:eq}).
Using (\ref{Fm:eq}) the DOS becomes
\begin{equation}
\varrho(E) = \sum_{{n=-\infty}\atop{\sigma=\pm 1}}^{\infty} 
\, \sum_{m=1}^{M_0} \, 
 \left| \frac{d I_{m,\sigma}(E)}{d E} \right| 
 \, \delta(I_{m,\sigma}(E) -n),
\label{rho-1}
\end{equation}
where $M_0 = M_{\sigma=+1}^{\left(- \right)}$, the number of propagating modes
for spin-up-hole.
Applying the Poisson summation formula\cite{Berry-1-Brack} 
to the summation over $m$  and keeping only the non-oscillating term 
one finds 
\begin{equation}
\varrho(E) = \sum_{{n=-\infty}\atop{\sigma=\pm 1}}^{\infty} \,
\left| \frac{d I_{m,\sigma}(E)}{d E} \right| 
%\rule[-1.6ex]{.2pt}{4ex}\;
\raisebox{-1.5ex}{$\scriptstyle {m=m^*}$}\, 
\frac{\Theta(M_0 - m^*)\Theta(m^*-1)}
{\left| \frac{\partial I_{m,\sigma}}{\partial m}\right|
%\rule[-1.6ex]{.2pt}{4ex}\;
\raisebox{-1.5ex}{$\scriptstyle {m=m^*}$}},
\label{rho-2:eq} 
\end{equation}
where the $E$-dependent $m^*$ satisfies Eq.~(\ref{Fm:eq}) for a given $n$
and $\sigma$. To simplify $I_{m,\sigma}$ we Taylor expand 
$k_{m,\sigma}^{\left(\pm\right)}$ in terms of $E+\sigma h $ 
in first order (which is consistent with the Andreev's approximation) 
and find
\begin{equation}
I_{m,\sigma}(E) = \left(E+\sigma h \right)\, 
\frac{2d/\left(\pi \hbar v_{\rm F}\right)}{\sqrt{1-{\left(m/M\right)}^2}} 
- \frac{\arccos E/\Delta}{\pi},
\label{Fm-approx:eq}
\end{equation}
where $v_{\rm F}$  is the Fermi velocity.  
In our approximation, $M_0\approx M$.  
From (\ref{Fm:eq}) and using (\ref{Fm-approx:eq}) we obtain
\begin{eqnarray}  
m^* &=& M \sqrt{1 - {\left(\frac{2d}{s_{n,\sigma}(E)}\right)}^2}, 
\,\,\, \text{where}  \nonumber \\[2ex] 
s_{n,\sigma}(E) &=&
\frac{n\pi +\arccos(E/\Delta)}{(E+\sigma h)/\Delta}\, \xi_0,
\label{sn:def}  
\end{eqnarray}
and $\xi_0 = \hbar v_{\rm F}/\Delta$ is the coherence length in the bulk 
superconductor. 

\begin{figure}[hbt]
\includegraphics[scale=0.4]{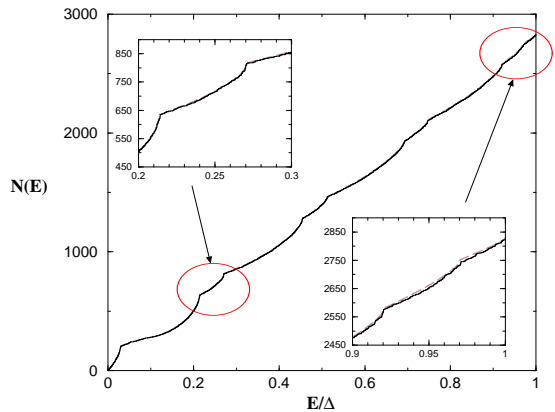}
\caption{Counting function $N(E)$ as a functions of $E/\Delta$
obtained from the exact (solid line) and the semi-classical calculation
(dashed line) for $h/\Delta=0.1$ ($h/E_F=0.0025$).  
The other parameters are $M=217.7$, $d/W=0.7$, $\Delta/E_{\rm
F}=0.025$.
The insets show the enlarged part of the main frame.
\label{N-verygood-fig}}
\end{figure}
\begin{figure}[hbt]
\includegraphics[scale=0.4]{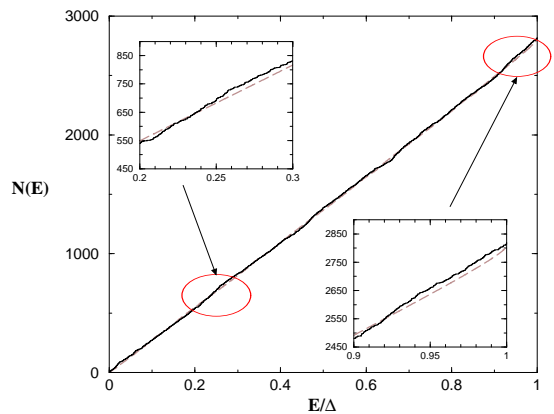}
\caption{Counting function $N(E)$ as a functions of $E/\Delta$
obtained from the exact (solid line) and the semi-classical calculation
(dashed line) for $h/\Delta=10.0$ ($h/E_F=0.25$). 
The other parameters are the same as in Fig.~\ref{N-verygood-fig}.
The insets show the enlarged part of the main frame.
\label{N-poor-fig}}
\end{figure}

Using (\ref{Fm-approx:eq}) and performing the derivatives 
in (\ref{rho-2:eq}) we find 
\begin{subequations}
\begin{equation}
\varrho(E) = \sum_{\sigma=\pm 1}
\frac{M}{\left|E+\sigma h\right|} \sum_{n=-\infty}^{\infty} \,
\left[ s_{n,\sigma}\left(E\right) + \xi_c \left(E\right) \right]
P(s_{n,\sigma}(E)),
\end{equation}
where
\begin{eqnarray}
P(s) &=& \frac{4 d^2}{s^3\sqrt{1-{\left(\frac{2d}{s}\right)}^2}}\, 
\Theta (s- 2d), 
\label{Ps-box}
\end{eqnarray}
\label{semi_DOS}
\end{subequations}
is a purely geometry-dependent function, 
$\xi_c \left(E\right) = \xi_0/\sqrt{1-E^2/\Delta^2}$ and 
$s_{n,\sigma}(E)$ is given by Eq.~(\ref{sn:def}). 
It can be shown that $P(s)$ is the classical probability 
that an electron entering the billiard at the FS interface 
returns to the interface  
after a path length $s$. The distribution $P(s)$ is normalized to
one, i.e., $\int_0^\infty\, P(s)\, ds =~1$.
Note that the result for the DOS, given in Eq.~(\ref{semi_DOS}) differs from 
that obtained in Ref.~\onlinecite{Nazarov-Ferro} by a factor multiplying 
$P(s)$ in the summation. 

\begin{figure}[htb]
\includegraphics[scale=0.4]{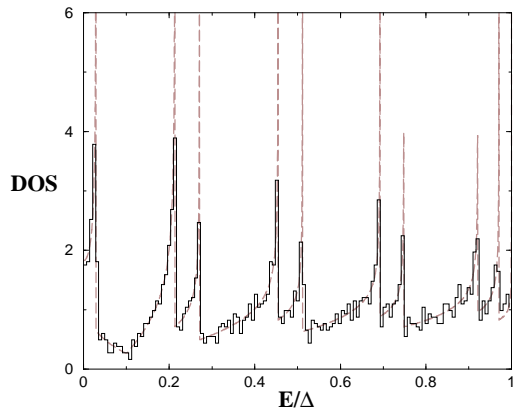}
\caption{The density of states $\varrho(E)$ as a functions of $E/\Delta$
(in units of $2\varrho_{\rm N}$) 
obtained from the exact (solid line) and the semi-classical calculation
(dashed line) for parameters given in Fig.~\ref{N-verygood-fig}. 
%$h/\Delta=0.1$ ($h/E_F=0.0025$), 
%$M=217.7$, $d/W=0.7$, $\Delta/E_{\rm F}=0.025$.
\label{dos-good-fig}}
\end{figure}
\begin{figure}[htb]
\includegraphics[scale=0.4]{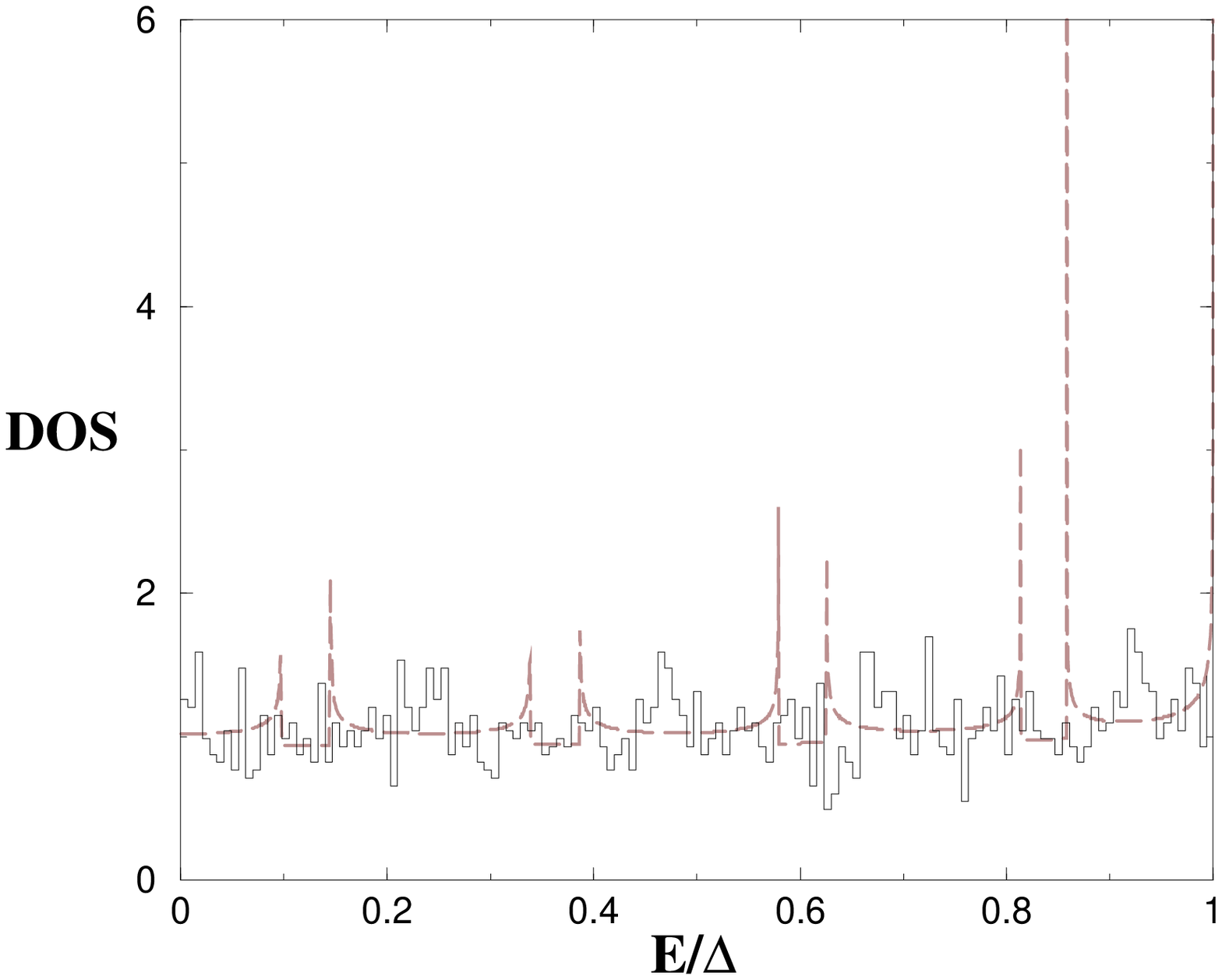}
\caption{The density of states $\varrho(E)$ as a functions of $E/\Delta$ 
(in units of $2\varrho_{\rm N}$) 
obtained from the exact (solid line) and the semi-classical calculation
(dashed line) for parameters given in Fig.~\ref{N-poor-fig}. 
%$h/\Delta=10.$ ($h/E_F=0.25$), 
%$M=217.7$, $d/W=0.7$, $\Delta/E_{\rm F}=0.025$.
\label{dos-poor-fig}}

\end{figure}
\begin{figure}[hbt]
\includegraphics[scale=0.4]{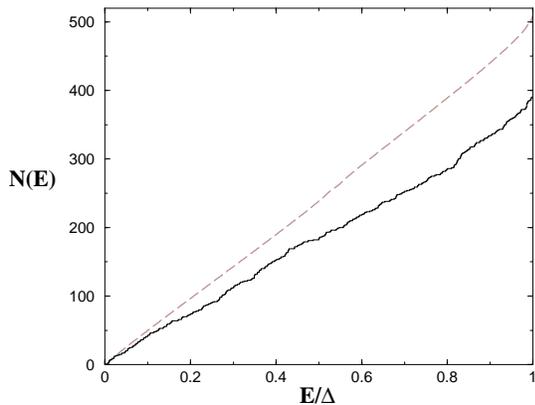}
\caption{Counting function $N(E)$ as a functions of $E/\Delta$
obtained from the exact (solid line) and the semi-classical calculation
(dashed line) for $h/\Delta=40.0$ ($h/E_F=1.0$). 
The other parameters are the same as in Fig.~\ref{N-verygood-fig}.
\label{N-breakdown-fig}}
\end{figure}

To compare the exact DOS obtained from the quantization condition 
(\ref{DNS}) with that calculated from the semi-classical 
expression (\ref{semi_DOS}), we introduce the integrated DOS: 
$N(E) = \int_0^{E}\, dE^\prime \varrho(E^\prime)$. From (\ref{semi_DOS}) we
have 
\begin{eqnarray}
N(E) &=& M \sum_{n=-\infty}^{+\infty} 
\left[ F_{n}^{\left(a\right)} (E) + F_{n}^{\left(b\right)} (E)\right], 
\text{where} 
\label{int_DOS} \\ 
F_{n}^{\left(a\right)} (E) &=&
F\left(s_{n,+1}(0)\right)-F\left(s_{n,+1}(E)\right),  \nonumber \\[1ex] 
F_{n}^{\left(b\right)} (E) &=&
\!\! \left\{\begin{array}{c}
1-F\left(s_{n,-1}(0)\right)-F\left(s_{n,-1}(E)\right), \,\, \text{if}\,\, E>h
\nonumber \\[1ex] 
-F\left(s_{n,-1}(0)\right)+F\left(s_{n,-1}(E)\right), \text{otherwise,}
\end{array}   \right. 
\end{eqnarray}
and $F(s)=\int_0^s \, P(s^\prime)ds^\prime 
=\Theta \left(s-2d \right)\, \sqrt{1-4d^2/s^2}$ is the integrated length
distribution. Note that $F(\infty)=1$ since $P(s)$ is normalized to 1.  

The small parameter used in reaching the semi-classical result is $h/E_F$. 
Figures \ref{N-verygood-fig} and \ref{N-poor-fig} show a comparison between 
the semi-classical result Eq.~(\ref{int_DOS}) and the exact result 
Eq.~(\ref{De}), when $h/E_F=0.0025$ and $0.25$, respectively.  
For better resolution, in the insets enlarged parts of the main
frames are shown.
Figures \ref{dos-good-fig} and \ref{dos-poor-fig} show the corresponding 
densities of states (in units of $2\varrho_{\rm N}$, where 
$\varrho_{\rm N}= m_{\rm F}A/(\pi \hbar^2)$ is the DOS of the ferromagnetic
region including spin up/down states and $A=Wd$ is the area of this region)  
with singularities given by the semi-classical formula 
$s_{n,\sigma}(E_{\text{sing}}) = 2d$, where $\sigma=\pm 1$ and 
$n$ is such that $E_{\text{sing}} < \Delta$. 
These figures show that the semi-classical formula given by 
Eqs.~(\ref{semi_DOS})-(\ref{int_DOS}) yield good agreement with the exact 
result, even for large values of $h/\Delta$, provided $h/E_F$ is
small. 
Figure~\ref{N-breakdown-fig} shows a comparison with the exact result 
when the latter condition is violated. 
Clearly the agreement is poor in this limit.

In summary we have shown that a semi-classical treatment of the clean limit 
yields an expression for the DOS in terms of the classical return 
probability $P(s)$, which in turn is known analytically. This formula is 
analogous to the quasi-classical result of Ref.~\onlinecite{Nazarov-Ferro}, 
where $P(s)$ is not known analytically and must be determined via a 
numerical simulation. We have also shown that the semi-classical
formula agrees very well with the exact result for small exchange 
field compared with the Fermi energy. 

We thank 
%C.~W.~J.~Beenakker and 
B.~L.~Gy{\"o}rffy for helpful discussions. 
This work was supported in part by the European Community's Human Potential
Programme under Contract No. HPRN-CT-2000-00144, Nanoscale Dynamics,
the Hungarian-British Intergovernmental Agreement on Cooperation in
Education, Culture, and Science and Technology, 
and the Hungarian  Science Foundation OTKA  TO34832.

%\bibliographystyle{prsty}
%\bibliography{/home/cserti/tex/cikkek.bib}

%\end{references}

\end{document}